\documentclass[fleqn,usenatbib]{mnras}

\usepackage{newtxtext,newtxmath}
\usepackage[T1]{fontenc}
\usepackage{ae,aecompl}
\usepackage{rotating}

\usepackage{graphicx,ulem}

\usepackage{bm}		
\usepackage[flushleft]{threeparttable}
\usepackage{float}
\usepackage[caption = false]{subfig}

\usepackage{adjustbox}
\usepackage{threeparttable}

\title[Radially-structured ionisation in AGN]{Steep X-ray reflection emissivity profiles in AGN as the result of radially-structured disc ionisation}

\author[E. S. Kammoun et al.]
       {E. S. Kammoun,$^{1,2\,{\thanks{E-mail: \href{mailto:ekammoun@umich.edu}{ekammoun@umich.edu}}\thanks{Former PhD student at SISSA.}}}$
        V.~Dom\v{c}ek,$^{3,4}$ J.~Svoboda,$^{5}$ M.~Dov\v{c}iak,$^{5}$ and G.~Matt$^{6}$\\
     $^1$Department of Astronomy, University of Michigan, 1085 South University Avenue, Ann Arbor, MI 48109-1107, USA\\
	$^2$SISSA, via Bonomea 265, I-34135 Trieste, Italy\\
	$^{3}$ Anton Pannekoek Institute / GRAPPA, University of Amsterdam, Science Park 904, 1098 XH Amsterdam, Netherlands \\
	$^{4}$ Department of Theoretical Physics and Astrophysics, Masaryk University, Kotl\'{a}\v{r}sk\'{a} 2, CZ-611 37 Brno, Czech Republic \\
	$^5$Astronomical Institute of the Academy of Sciences, Bo\v{c}n\'{i} II 1401, CZ-14100 Prague, Czech Republic\\
	$^6$Dipartimento di Matematica e Fisica, Universit\`{a} degli Studi Roma Tre, via della Vasca Navale 84, 00146 Roma, Italy\\
	}

\date{Accepted 2018 Month Day;
      Received 2018 Month Day;
      in original form 2018 Month Day}
\pubyear{2018}

\begin{document}
\label{firstpage}
\pagerange{\pageref{firstpage}--\pageref{lastpage}}
\maketitle

\begin{abstract}
X-ray observations suggest high compactness of coron\ae\ in active galactic nuclei as well as in X-ray binaries. The compactness of the source implies a strong radial dependence in the illumination of the accretion disc. This will, for any reasonable radial profile of the density, lead to a radial profile of the disc ionisation. \cite{2012A&A...545A.106S} showed on a single example that assuming a radially-structured ionisation profile of the disc can cause an artificial increase of the radial-emissivity parameter. We further investigate how the X-ray spectra are modified and quantify this effect for a wide range of parameters.  Computations are carried out with the current state-of-the-art models for relativistic reflection.  We simulated spectra using the response files of the micro-calorimeter X-IFU, which is planned to be on board of Athena. We assumed typical parameters for X-ray bright Seyfert-1 galaxies and considered two scenarios for the disc ionisation: 1) a radial profile for the disc ionisation, 2) a constant disc ionisation. We found that steep emissivity profiles can be indeed achieved due to the radial profile of the disc ionisation, which becomes more important for the cases where the corona is located at low heights above the black hole and this effect may be even more prominent than the geometrical effects. We also found that the cases with high inner disc ionisation, rapidly decreasing with radius, may result in an inaccurate black hole spin measurements.
\end{abstract}


\section{Introduction}\label{intro}

\begin{figure}
\includegraphics[width=0.99\linewidth]{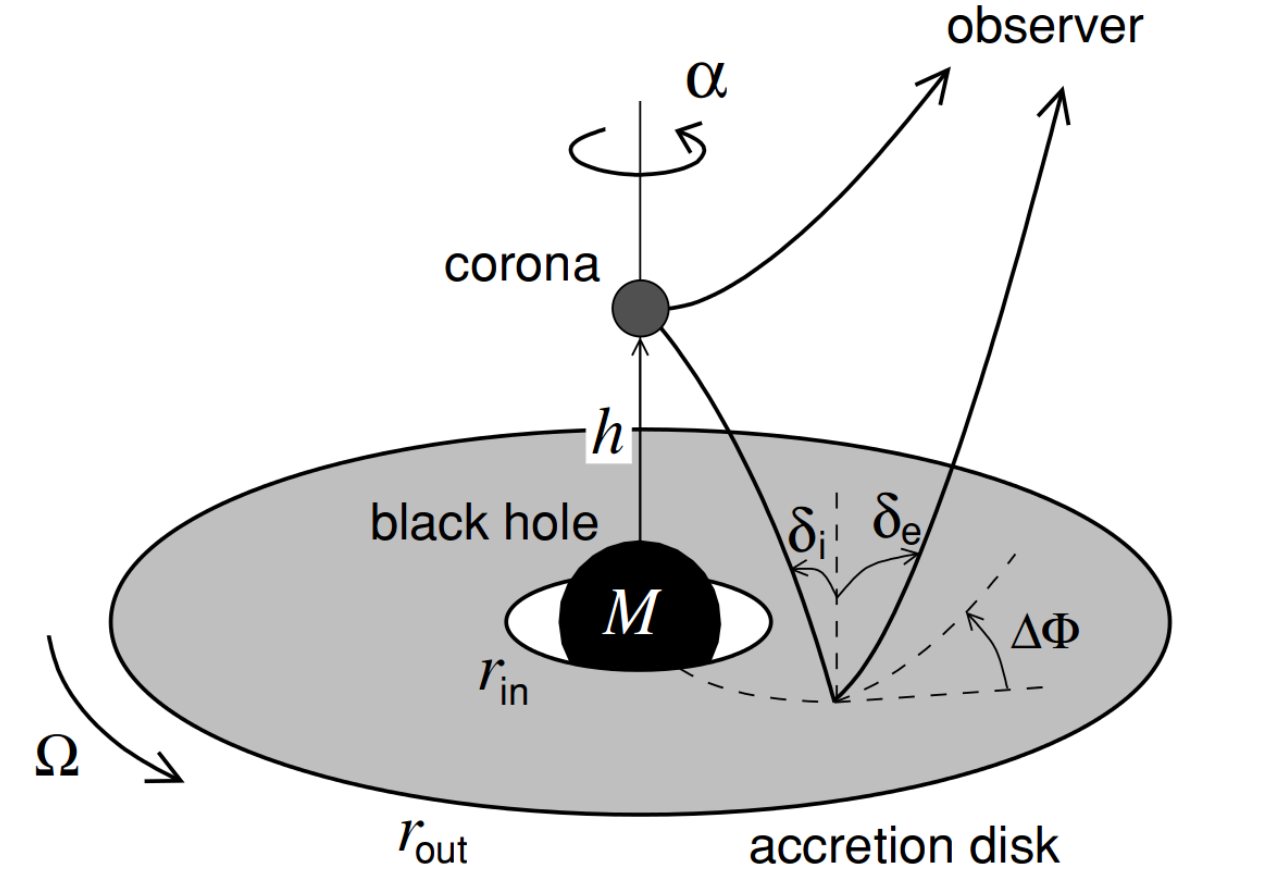}
\centering
\caption{Lamp-post geometry scheme. Figure adapted from \citet{2011ApJ...731...75D}.
}

\label{lamp-post}
\end{figure}

X-ray spectroscopy of active galactic nuclei (AGNs) and X-ray binaries (XRBs) provides a good opportunity to study the physics of accretion on supermassive black holes \citep[e.g.][and references therein]{Brandt2015, Merloni2016}. The X-ray emission originates in a hot medium (so called `corona') located above the accretion disc consisting of relativistic electrons that inverse-comptonise the thermal ultraviolet light, emitted by the accretion disc. The X-ray emission from the corona is partly emitted in the direction of the observer (known as `primary spectrum') and partly in the direction of the accretion disc, as shown in Fig.~\ref{lamp-post}. In the latter case, the light will be reprocessed by the accretion disc and re-emitted towards the observer (known as `reflection spectrum'). The evidence for such reflected emission from the accretion disc is the presence of broad fluorescent iron K$\alpha$ lines (at $\sim 6.4$~keV), in addition to a broad hard X-ray excess peaking at 20-30~keV (known as Compton hump) reported in X-ray spectra of several AGN  \citep[e.g.][]{1995Natur.375..659T, Walton2012, Marinucci2014a, Parker2014, Svoboda2015}. 

Measuring the properties of X-ray radiation from the innermost regions allows us to constrain the properties of the corona and gives a potential way to estimate black hole spins \citep[for a review see, e.g.,][and references therein]{Reynolds2013}. AGN X-ray spectra  showing broad iron lines need therefore to be fitted with models that account for the relativistic effects, such as the gravitational and Doppler frequency shifts, light bending, aberration, and Doppler boosting. Several numerical codes have been developed in order to estimate these relativistic effects on the X-ray reflection spectra and can be used for spectral fitting \citep[e.g.][]{ky, 2004MNRAS.352..353B, 2006ApJ...652.1028B, Niedzwiecki2008, Dauser2010, Nied2016}. The ultimate goal of such analyses is to measure the black hole spin. However, such measurements are strongly affected by a mutual degeneration of the model parameters. The spin measurements depend on the inclination of the system \citep{ky, 2004MNRAS.352..353B}, but also on parameters that account for the geometry of the corona, namely the X-ray reflection emissivity profiles \citep[see, e.g.,][]{2004MNRAS.352..353B, Svoboda2009, 2012A&A...545A.106S}, reflection fractions \citep{Dauser2014} or directly the height of the corona \citep{Dovciak2014, Kammoun18}, assuming the so called lamp-post geometry as shown in Fig.~\ref{lamp-post}.

The lamp-post geometry is a simplification of a compact centrally-located corona that has been suggested  mainly from the measurements of a steep radial emissivity of X-ray spectra of several AGNs and XRBs. The emissivity index $q$ describes how much the reflected emission decreases with the radius (see eq.~\ref{qdef}). Steep indices were commonly measured in Seyfert 1 galaxies. For example, \cite{Walton2013} showed that 15/25 bare AGNs, from their sample, reveal steep emissivities, larger than 5 (see their tables 2 and 3). Five other cases in their samples had emissivities indices steeper than 4, while they could not constrain the emissivity profiles for rest of the sources so they fixed it to 3. More recent measures have also shown steep emissivities in AGNs, for example, 1H0707-495 \citep[$q_{\rm in} = 10.2_{-0.9}^{+0.8}$, $q > 8.6$,  $q=7.3_{-0.2}^{+0.4}$;][respectively]{2012MNRAS.422.1914D, Fabian20121H, Kosec18}, IRAS\,13224-3809 \citep[$q_{\rm in} = 7.6_{-0.2}^{+0.5}$;]{Jiang18}, Mrk 335 \citep[$q_{\rm in} \sim7.7-9.9$;][]{Wilkins2015} or Swift J2127.4+5654 \citep[$q = 6.3_{-1.0}^{+1.1}$;][]{Marinucci2014}. The steep emissivity in X-ray reflection spectra was also found in several cases of black hole X-ray binaries, such as Cyg\,X-1 \citep[$q_{\rm in} > 7$;][]{2012MNRAS.424..217F, Tomsick2014}, GRS 1915+105 \citep[$q_{\rm in} > 7$;][]{Miller2013}, MAXI J1535-571 \citep[$q_{\rm in} = 7.8$;][]{2018ApJ...860L..28M}, XTE\,J1650-500 \citep[$q \approx 4-5$;][]{2004MNRAS.351..466M}, or GX\,339-4 \citep[$q \approx 6$;][]{2007ARA&A..45..441M}.

The measured indices often reach values up to $q \approx 7-10$, which can be hardly explained without requiring an extremely compact X-ray source \citep{2012A&A...545A.106S, Dauser2013, Dovciak2014}, or an extended corona but at a very low height above the inner disc \citep[][see Section \ref{sec:emissivityprofile} for further discussion regarding the various emissivity laws and geometries.]{Wilkins2011}. Another independent indication of the compactness of the corona comes from the gravitational microlensing of quasars. Recent results from monitoring observations of lensed quasars showed that the X-ray corona can indeed be confined within a half-light radius of $\sim 6\, \rm r_g$ \citep{Chartas09, Mosquera13, Reis13}. The compact X-ray source has to cause a significant radial profile of the disc irradiation. Consequently, when any realistic profile of the density is considered, the ionisation also significantly decreases with the radius \citep{2012A&A...545A.106S}. This effect is, however, totally neglected in current spectral fittings, which assume a constant ionsation of the accretion disc that is fitted as an independent parameter. This may then break the self-consistency of the models, thus leading to inaccurate physical interpretations. \citet{2012A&A...545A.106S} showed on a simple example that ignoring this effect may indeed cause measurements of artificially steeper radial emissivity indices for a case of highly compact corona located at low height, assuming a lamp-post geometry. In this work, we significantly extend the previous analysis by performing a systematic analysis of this effect. For this reason we consider different configurations of lamp-post height, disc ionisation and disc density. We further investigate in more details how these factors may affect the measurements of emissivity profiles as well as the spin parameter.

\section{Model assumptions}\label{model}

\subsection{Lamp-post scheme of the corona and radial emissivity profile}
\label{sec:emissivityprofile}

The lamp-post scheme of the corona \citep{George89, Matt1991}, as shown by a schematic illustration in Fig.~\ref{lamp-post}, represents the simplest geometry of a compact X-ray source. The idea of a compact X-ray source positioned on the rotational axis of the black hole was also used to explain the observed X-ray variability of AGN \citep{Miniutti2004, Niedzwiecki2010}. The flux decreases when the height of the corona decreases because of the larger photon capture by the central black hole. Moreover, at low heights, more primary radiation will be focused towards the innermost regions of the accretion disc due to light-bending and gravitational redshift which decreases the reflected flux as well \citep[see e.g.][]{Martocchia96}. The reflection fraction therefore increases in this state \citep[for quantitative estimates see, e.g.,][]{Dauser2013}.

However, in a more realistic situation the corona may be a more complex inhomogeneous medium extended in both radial and vertical direction to larger radii \citep[e.g.][]{Wilkins2015}. \citet{Dovciak2016} pointed out that the source needs to be extended to be able to produce sufficiently enough X-ray photons to match the observations. However, the more complex models that would account for the spatial extension of the corona would contain more free parameters that would be difficult to be uniquely constrained with the current quality of the data. Therefore, owing to its simplicity, the lamp-post scheme is still popular and frequently used in the most recent codes for relativistic smearing \citep{Dauser2013, Dovciak2014}. Besides to an isotropic homogeneous corona, the lamp-post scheme represents a simple approximation of a spatially compact corona, which is concentrated towards the centre.

Often, instead of assuming any particular geometry, the radial emissivity profile, with an emissivity index $q$, was introduced in the relativistic reflection models:
\begin{equation}
\label{qdef}
\epsilon (r) \propto r^{-q},
\end{equation}
\noindent 
For an isotropic corona, the thermal energy dissipation is assumed to decrease with the third power of radius ($q=3$), following the standard prescription of the accretion disc temperature \citep{1973A&A....24..337S}. Also, for the lamp-post geometry, the irradiation at distant parts of the accretion disc should follow $r^{-3}$ \citep{1997ApJ...488..109R}. Thus, $q=3$ is considered as a `standard' index of the emissivity. However, the emissivity profile in the lamp-post geometry significantly changes in the innermost regions depending on the location of the X-ray source \citep[see e.g.][]{Martocchia00, Martocchia02,Wilkins2011, Dauser2013, Dovciak2014}. Assuming an accretion disc with a constant density and ionisation, $q$ can reach large values only when the source is located very close to the black hole event horizon. In this case, the emissivity profile is very steep at the innermost radii, then it flattens and finally reaches $q=3$ at further radii, where the contribution to the total reflection spectrum is often relatively small. Therefore, with the current quality of the data, broken (or twice-broken) power laws can be used as adequate approximations of the intrinsic emissivity profiles \citep[e.g.][]{Wilkins2012, Gonzalez17}.

\subsection{Disc irradiation and the ionisation profile}

\begin{figure}
\centering
\includegraphics[width=0.49\textwidth]{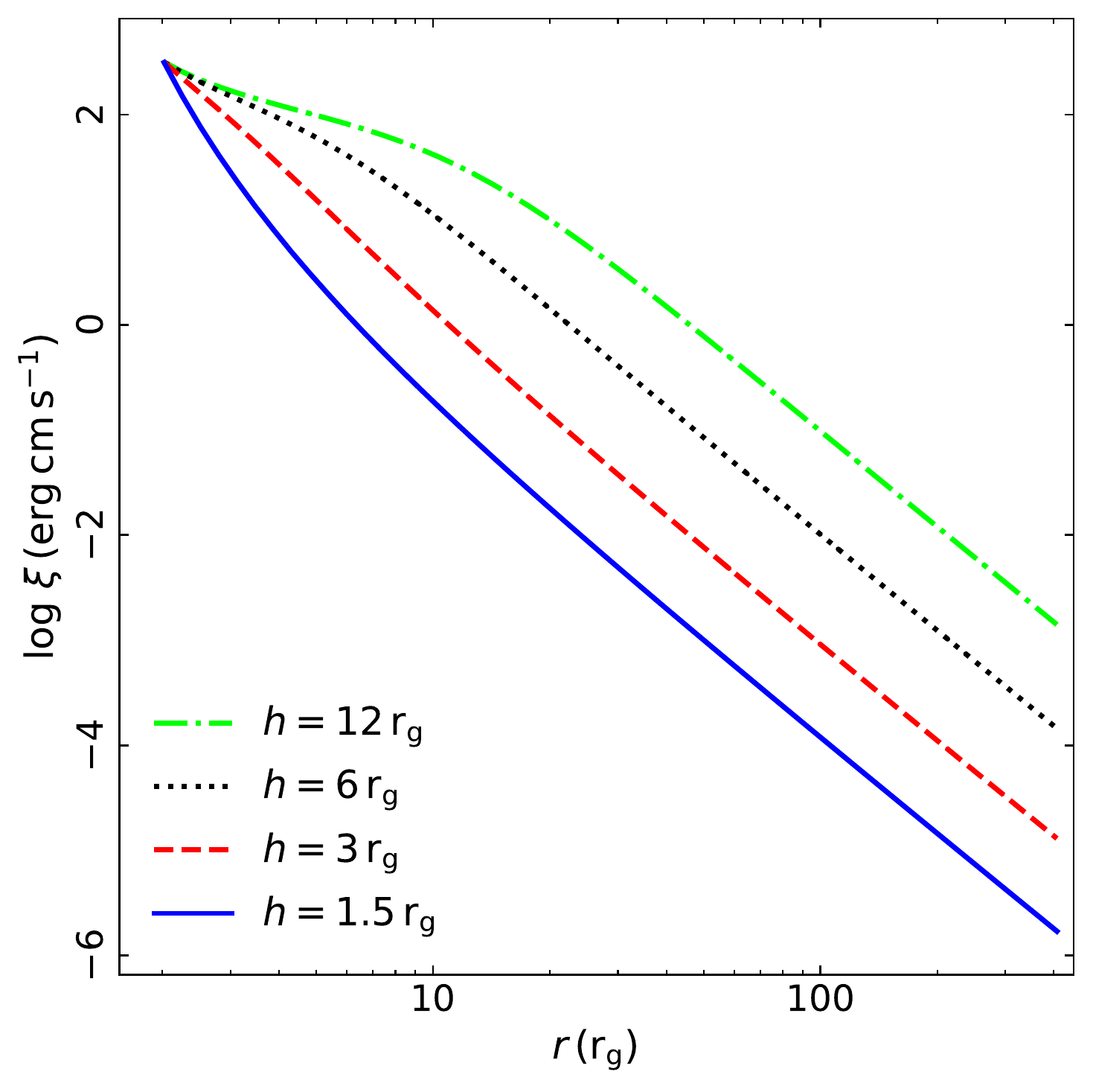}
\caption{The radial ionisation profile $\xi(r)$, estimated using the {\tt KYNXILLVER} model, for various lamp-post heights 1.5~$\rm r_g$ (solid blue line), 3~$\rm r_g$ (dashed red line), 6~$\rm r_g$ (dotted black line), and 12~$\rm r_g$ (dash-dotted green line). We assume the same inner ionisation parameter ($\log \xi_{\rm in} = 2.5$) for all the cases.}
\label{ionisation_fig}
\end{figure}

The accretion disc material, irradiated by the X-ray photons of the corona, will be then naturally ionised, depending on the illuminating flux. The strong radial dependence of the irradiation, discussed in the previous section, will consequently lead to a radial dependence in terms of ionisation as well. Therefore, the ionisation parameter ($\xi$) at each radius, $r$, of the disc can be described as:

\begin{equation}
\xi (r) = \frac{4\pi F_{\rm inc}(r)}{ n_{\rm H } },
\end{equation}
\noindent 
where $F_{\rm inc}$ is the irradiating flux and $n_{\rm H}$ is the density of the disc, assumed to be constant. Fig.~\ref{ionisation_fig} shows the dependence of the ionisation profile on the height of the lamp-post. By increasing the height of the lamp-post further out regions of the disc are illuminated, which leads to a flattening in the ionisation profile below $\sim 20~\rm r_g$. In the radial extension, this assumption is satisfied because the radial dependence of the irradiation is much stronger than the radial dependence of any realistic density profile \citep[see Fig.~3 in][]{2012A&A...545A.106S}.


The vertical structure of the disc would cause that the reflection at each radius is composed of reflection from different layers and thus with different ionisations \citep[e.g.][]{Zycki1994, Nayakshin2000, Ballantyne2001}. However, we simply assume that more ionised emission would still more likely come from the innermost, regions where the irradiation is much higher, due to light bending. Thus, the decrease of the reflection emission with radius, for an average ionisation level, should be preserved similarly to the case without any vertical structure. Considering the vertical stratification of the disc density would then require employing more complex models that are not currently available for the fast spectral fitting required in this analysis. Therefore, we potentially left studying the effect of vertical stratification for future investigations.

In this work, we used the {\tt XILLVERD} \citep{Garcia16} tables for the ionised reflection model. This model assumes a power-law primary illuminating an accretion disc for a given ionization parameter and density.  We note that the effect of ionisation is the most prominent below $\sim 10\,\rm keV$. For the cold/neutral disc  (low $\xi_{\rm d} $), a number of emission lines are expected in the soft X-rays. As the disc gets warmer and more ionised, the equivalent width of the emission lines decreases. In the most ionised case, no lines originate by reflection from such ionised plasma. We note that different ionisation levels also influence the shape of the reflection spectrum. While the highly-ionised reflection spectra resemble a power-law shape decreasing with the energy, the cold reflection spectra have less emission in the soft band and a more prominent Compton hump. Consequently, there is a lower spectral power-law index when comparing the cold and highly ionised reflection. We also note that the larger the disc density the larger the soft flux (below $\sim 1$~keV). This is attributed to an increase in the surface temperature of the disc as the density increases \citep[see][for more details about the effect of disc ionisation and density on the reflection sepctrum]{Garcia2013, Garcia16}.



\subsection{Relativistic reflection combining the lamp-post geometry and radially-structured ionised reflection}
\label{seed_model}

By combining the relativistic smearing code from the {\tt KYN} package \citep{Dovciak2004} and the ionised reflection model, we get the resulting model {\tt KYNXILLVER}\footnote{\url{https://projects.asu.cas.cz/stronggravity/kyn\#kynxillver}}, which incorporates both the lamp-post geometry \citep{Dovciak2014} and the radially-structured ionisation, self-consistently calculated from the X-ray irradiation.  It also includes the primary power law continuum that goes directly to the observer. The model can be described as:
\begin{equation}
{\tt KYNXILLVER} = {\tt powerlaw} + \sum {\tt KYNCONV} (\Delta r_i) * {\tt XILLVER}(\xi _i).
\label{kyreflionx_mo}
\end{equation}

\noindent It represents the total X-ray emission of the source and reflection from the accretion disc with different ionisation level in each annulus. The primary flux is assumed to be isotropic in the local frame of the lamp-post. In other words, the amount of radiation emitted, per solid angle, in the direction of the observer is equal to the one emitted in the direction of the disc is the same. This can be done by setting the {\tt KYNXILLVER} parameter ${\tt Np:Nr =1}$. The model then accounts self-consistently for the relativistic effects acting on the primary radiation, depending on the height of the source. 

\section{Methods}
\label{sec:method}

\begin{table}
\centering
\caption{Key parameters used to perform the simulations with the corresponding input range.}
\begin{tabular}{lc}
\hline
Parameter	&	Input value	\\ \hline \\[-0.2cm]
$a^\ast$	&	0.94	\\[0.2cm]
$M\, (\rm M_{\odot})$	&	$10^7$	\\[0.2cm]
$\theta$ (deg.)	&	30	\\[0.2cm]
$r_{\rm in}\,\rm (r_{ISCO})$	&	1	\\[0.2cm]
$r_{\rm out}\,\rm (r_g)$	&	400	\\[0.2cm]
$\log \left(\frac{ \xi_{\rm in}}{\rm erg~cm~s^{-1}}\right)$	&	0--4.5	\\[0.2cm]
$h\,(\rm r_g)$	&	[1.5, 3, 6, 12]	\\[0.2cm]
$\Gamma$	&	2	\\[0.2cm]
$A_{\rm Fe}$ (solar)	&	1	\\[0.2cm]
$n_{\rm H}\,\rm (cm^{-3})$	&	$10^{15-19}$\\	\hline

\end{tabular}
\label{table:input_model}
\end{table}

\subsection{Data simulation}
\label{data}

\begin{figure*}
\centering
\includegraphics[width=0.99\linewidth]{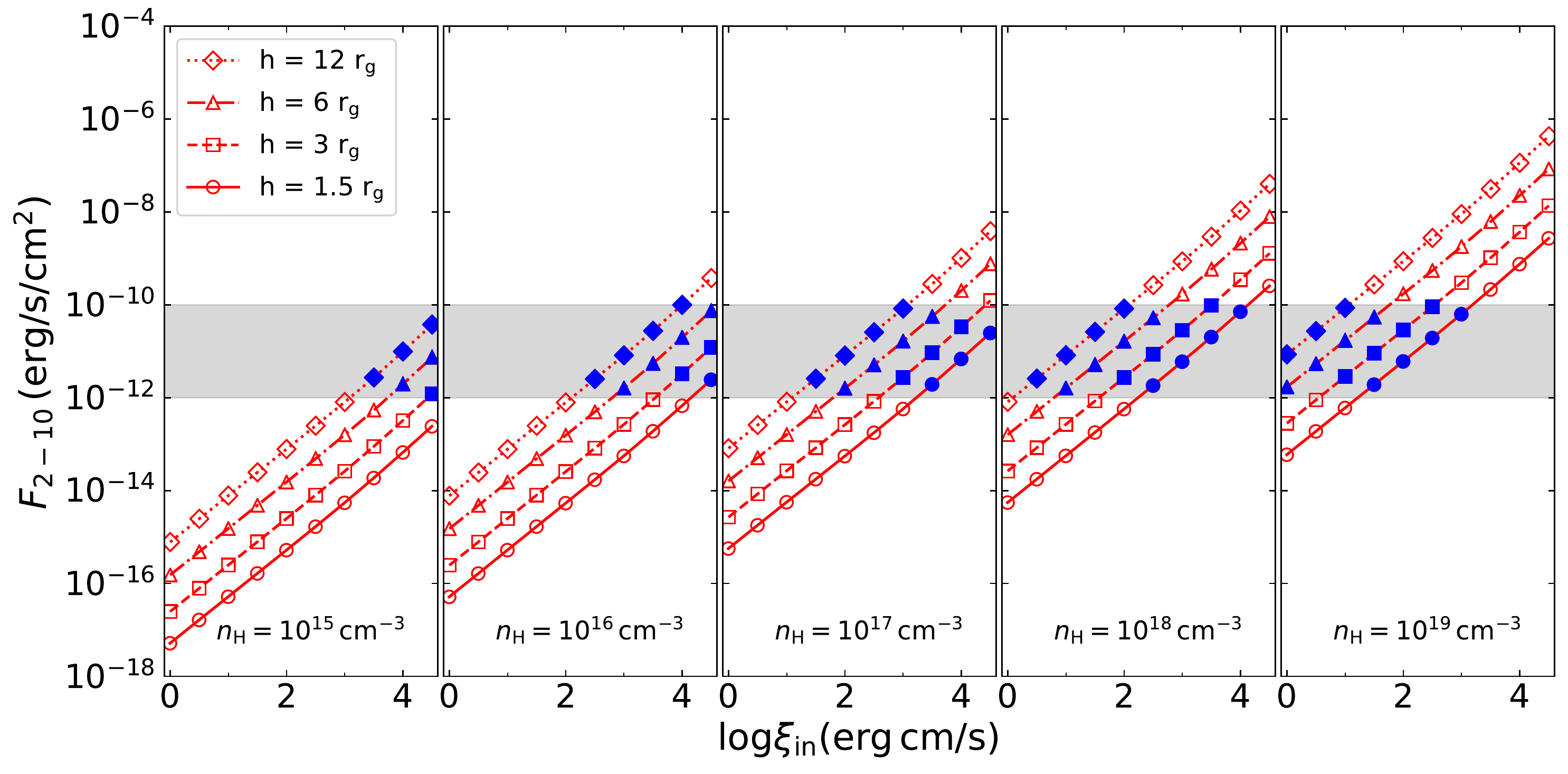}
\caption{Observed 2--10~keV flux as function of the ionisation parameter at the ISCO, for the various values of $n_{\rm H}$ and $h$ which are considered for this work. The shaded regions represent the flux range of $10^{-12}-10^{-10} ~\rm erg ~ s^{-1} ~ cm^{-2}$. The blue points represent the cases that are simulated in this work (see Section ~\ref{data} for more details.).}
\label{flux_fig}
\end{figure*}

The spectral simulations and the consequent analysis, were perfomed using a combination of the software tools for X-ray spectral analysis \MakeUppercase{\tt Xspec~v12.10} \citep{Arnaud1996} and Python as the \MakeUppercase{\tt Xspec}'s wrapper and plotting device. First we created theoretical spectra assuming the following seed model (in the XSPEC terminology),
\begin{equation} \label{eq:model_seed} 
{\tt model_{seed} =  phabs \times KYNxillver},
\end{equation}
\noindent where the {\tt phabs} model accounts for photo-electric absorption which is present in the line-of-sight of our Galaxy (we assume a column density $N_{\rm H} = 4 \times 10^{20}\,\rm cm^{-2}$). Throughout our analysis, we used values of typical X-ray bright Seyfert 1 galaxies, presented in Table~\ref{table:input_model}. We note that the parameters of the {\tt KYNXILLVER} model which are not presented in this table are fixed to their default values. We considered five values for the disc density in the range $10^{15}-10^{19}\,\rm cm^{-3}$, four lamp-post heights $h=1.5,~3,~ 6,~ 12\, \rm r_g$. We modified the intrinsic luminosities of the primary source as defined by the {\tt KYN} model in a way that, for each case, the inner value of the ionisation parameter at the innermost stable circular orbit (ISCO) $\log \xi_{\rm in}$ is between 0 and 4.5 (we considered 10 values of $\log \xi_{\rm in}$ spaced by a step of 0.5). Fig.~\ref{flux_fig} shows the total observed 2--10~keV flux for each of the considered cases. Our aim is to investigate how the radial-emissivity parameter changes in response to different heights of the corona and different densities of the accretion disc, by simulating spectra of bright local AGN. For that reason we chose only the cases that represent a flux level between $10^{-12}$ and $10^{-10}~ \rm erg~s^{-1}~cm^{-2}$, plotted as blue points in Fig.~\ref{flux_fig}. We note that this figure clearly shows the restrictions in the parameter space which are present in flux-limited samples. For example, for low density ($n_{\rm H} = 10^{15}~\rm cm^{-3}$) and low height ($h = 1.5~\rm r_g$), the intrinsic luminosity has to be small in order to have accretion discs with mild ionisation ($\log \xi_{\rm in} = 2$, for example). This would result in low observed flux $\sim 10^{-15}\rm erg\,s^{-1}\,cm^{-2}$. Such sources, if they exist, cannot be detected with reasonable exposures. Studying various configurations and their observational implications are beyond the scope of this work and left for future investigations.

Then, we used the {\tt XSPEC} command {\tt fakeit} to create fake spectra, based on the blue points in Fig. \ref{flux_fig}, considering the response files of the X-ray Integral Field Unit \citep[X-IFU;][]{Ravera14} which will be on board of the planned ESA mission Athena \citep{Nandra2013}. We assumed exposure times of 100~ks, with no background files. The spectral analysis throughout this work is performed using the Cash statistics \citep[$C-$stat;][]{Cash76} without rebinning neither the spectra nor the response files. 



\subsection{Spectral analysis model}
The data were created using a complex model with variable radially-structured ionisation of the disc. However, for the spectral analysis we used a simpler model (in the XSPEC terminology):

\begin{align} \label{eq:model_fit}
{\tt model_{fit}}	& {\tt =  phabs \times (KYconv * atable \{ xillverD-4.fits \}  } \nonumber \\
 						& {\tt + powerlaw )} ,
\end{align}
\noindent
where {\tt KYconv} is a convolution model which we apply to the {\tt XILLVERD} reflection tables assuming a single power-law radial emissivity and a constant ionisation of the disc, and the {\tt powerlaw} component represent the primary emission. Such model configuration has been standardly used in the X-ray data analysis when relativistic reflection is taken into account \citep[e.g.][]{2010MNRAS.406.2591P, Marinucci2014, Kosec18}, and we aim to investigate how the emissivity index is affected if the intrinsic data are more complex with radially-structured ionisation. 

Broken or twice-broken power-law emissivity profiles have also been used for X-ray spectral fitting with the relativistic reflection models since they better correspond to the emissivity profile of the source in the lamp-post geometry \citep{Wilkins2011,Wilkins14}. The emissivity is very steep at the innermost region, then flattens and finally it follows $r^{-3}$ profile. However, these more complex emissivity profiles may lead to a degeneracy between the emissivity indices and the break radii. Thus, assuming a single power-law emissivity profile diminishes a possibility that the best-fit would result in a steep emissivity index but in a very tiny area of the innermost region, and it is appropriate enough to understand the effect of the radial disc-ionisation profile on the measured emissivity indices.


\begin{figure}
\centering
\includegraphics[width=0.9\linewidth]{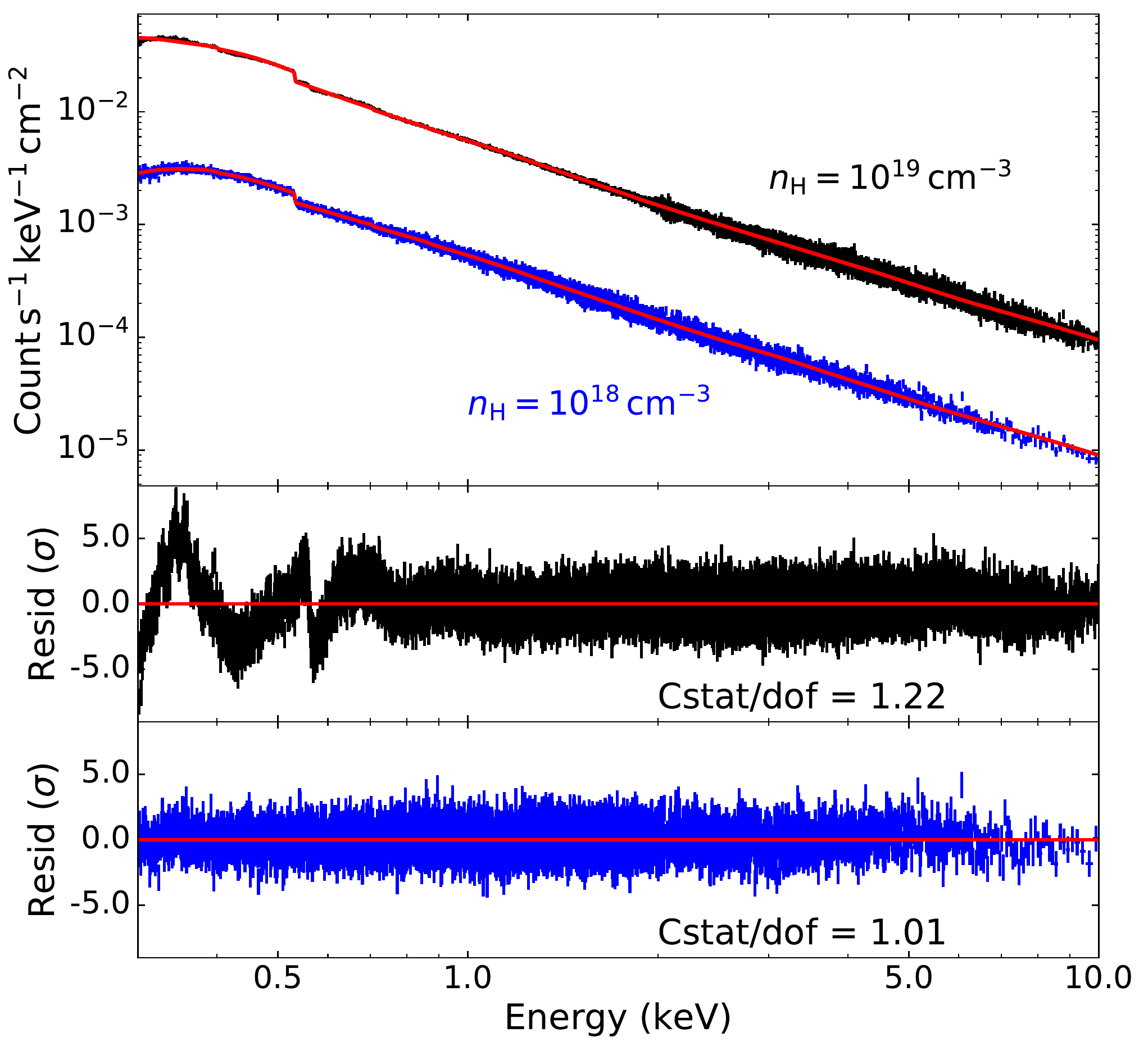}
\caption{Simulated spectra assuming $\log \xi_{\rm in} = 2.5$, $h=1.5~\rm r_g$ and densities $n_{\rm H} = 10^{18}~ \rm cm^{-3}$ (blue points) and $n_{\rm H} = 10^{19}~ \rm cm^{-3}$ (black points) together with the correspondant best-fit models (red lines). The residuals for the latter and former spectra are plotted in the bottom and middle panels, respectively. The spectra are binned for clarity.}
\label{spectra_fig}
\end{figure}

\begin{figure*}
\centering
\includegraphics[width=0.99\linewidth]{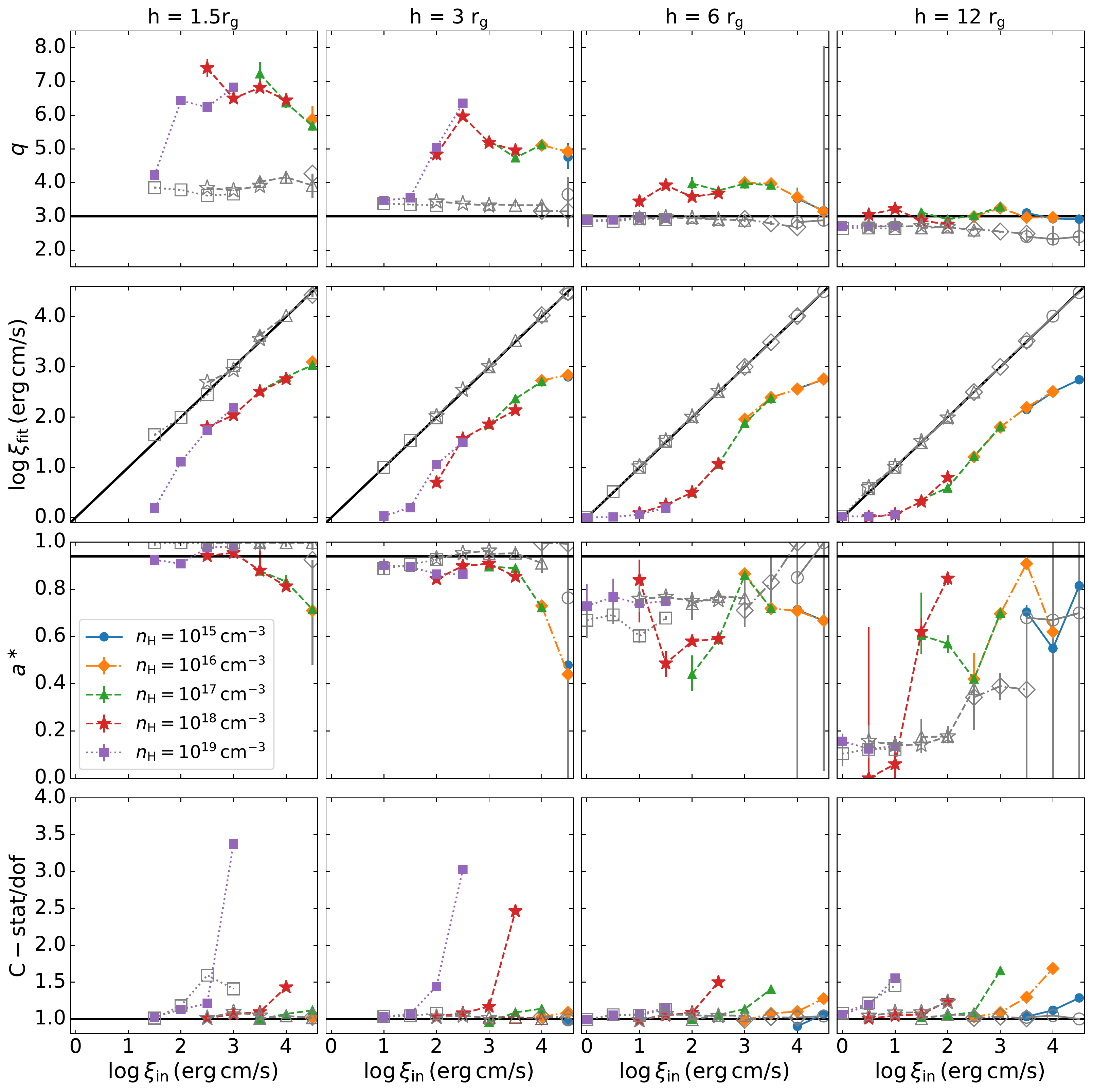}
\caption{Best-fit parameters, emissivity index, ionisation parameter, spin parameter, goodness of fit (from top to bottom rows, respectively), obtained by fitting the simulated spectra, for the various lamp-post heights and disc densities that we consider in this work. The simulations were performed assuming a radial ionisation profile (filled symbols) and constant ionisation (empty symbols) for the accretion disc. The black solid lines represent: the identity line (upper row), the Newtonian limit for the emissivity profile ($q=3$), the input value of the spin parameter ($a^\ast = 0.94$) and $\rm C-stat/dof =1$ (top to bottom rows, respectively; See Section \ref{sec:results} for more details).}
\label{Par_fig}
\end{figure*}

Our model has five free parameters, namely: the spin ($a^\ast$), the emissivity index ($q$), the ionisation of the disc ($\xi_{\rm fit}$), and the normalizations of the power-law ($N_{\rm PL}$) and the reflection ($N_{\rm XILLVER}$) components. The other parameters of the model were fixed to their input values. The fits were statistically accepted in most of the cases, except for the cases were the fluxes are very high $F_{2-10}\sim 10^{-10}\,\rm erg~ s^{-1}~cm^{-2}$, where some residuals could be seen below $\sim 0.8$~keV. We show in Fig.~\ref{spectra_fig} two examples of the simulated spectra together with the corresponding best-fit models. These spectra were simulated assuming  $\log \xi_{\rm in} = 2.5$ and densities $n_{\rm H} = 10^{18}~ \rm cm^{-3}$ and $n_{\rm H} = 10^{19}~ \rm cm^{-3}$, which according to Fig.\,\ref{flux_fig} correspond to different flux level. The residuals at soft X-rays for the bright case ( $n_{\rm H} = 10^{19}~ \rm cm^{-3}$) can be seen clearly in the middle panel of Fig.~\ref{spectra_fig}, while the residuals for the low-flux case can be well acceptable (see bottom panel of the same figure). This clearly demonstrates that the radially-structured ionisation can be easily undetectable in most of the data analysis. We further study how the parameters of the reflection model could change to mimic the intrinsic reflection emission with radially-structured ionisation.

\section{Results}
\label{sec:results}

The results obtained by fitting the simulated data with the model presented in eq. \eqref{eq:model_fit} are shown in Fig.\,\ref{Par_fig}. The $x$-axis represents the inner ionisation parameter that was used in the simulations. The plots show (from top to bottom) the best-fit values of the emissivity index, ionisation parameter of the constant-ionisation model, spin and the goodness of the fit ($C$-stat/dof). Different heights were considered in the simulations ($h = 1.5, 3, 6$ and 12 $\rm r_g$ from left to right). The coloured (filled) symbols represent the cases for different accretion-disc densities and the radially structured ionisation according to the lamp-post source illumination.

The emissivity index \textit{q} increases up to $\sim 7$ and $\sim 5$ for heights of 1.5 and 3 $\rm r_g$, respectively, while it approaches the limit of $q \sim 3$ for larger heights. In fact, $q$ is low for $\log \xi_{\rm in} < 1.5$ and it increases with larger $\xi_{\rm in}$, then it again decreases for high values of $\log \xi_ {\rm in} > 4$. This effect is an immediate consequence of the ionisation gradient of the disc (shown in Fig.\,\ref{ionisation_fig}). For low and large values of $\xi_{\rm in}$, the parts of the disc contributing the most to the observed spectrum would be either neutral or highly ionised, depending on the height of the lamp-post. Thus, a very small gradient of ionisation would be expected in these parts of the disc, and the constant-ionisation approximation holds in these extreme cases, which apparently results in lower values of $q$. However, for the intermediate values of $\xi_{\rm in}$, the gradient of ionisation is more important. The innermost regions will be more ionised and will have softer reflection spectra with respect to further out regions of the disc which are less ionised with a non-negligible contribution to the overall spectrum. Thus, a model assuming a single ionisation parameter of the disc will underestimate the ionisation from the innermost regions, by assuming an average ionisation. This effect will be then compensated by requiring a steep emissivity profile (see Fig.4 of \citealt{2012A&A...545A.106S}, for more details about how the reflection spectra look like for different radii with the radially decreasing ionisation).

We note that the disc density plays a minor role in determining the emissivity profile. The values of $q$ for different values of $n_{\rm H}$, assuming the same $\xi_{\rm in}$, are consistent with each other. This leave us with two interplaying factors which may affect the emissivity profile: a) the geometry (height of the lamp-post) and b) the ionisation gradient. In order to investigate which of the two factors, or if it is the combination of both, that is leading to steep emissivity indices we performed the following experiment. We followed the same procedure described in Section \ref{sec:method} but considering a constant ionisation for the disc. In this case, both the simulations and the spectral fits are performed assuming a disc with a constant ionisation. This will allow us to evaluate quantitatively how much the geometry of the source would play a role in giving rise to steep emissivity profiles. The best-fit parameters are shown as open symbols in Fig.~\ref{Par_fig}. The emissivity indices are almost constant for the various ionisation states, assuming the same lamp-post height. The average values of $q$ are 3.89, 3.35, 2.89, and 2.59 for $h = 1.5$, 3, 6, and 12$~\rm r_g$, respectively. This clearly indicates that the steepening in emissivity profiles that can be introduced {\it only}  by the geometry of the corona is much smaller than the one introduced by considering the radial profile of the disc ionisation as well.

The third key parameter in our analysis is the black hole spin. It is clear from the lower panel of Fig.~\ref{Par_fig} that the correct value of $a^\ast = 0.94$ can be recovered for the cases of low heights (1.5 and 3~$\rm r_g$). While for larger heights, the spectral fits suggest lower and unconstrained spin values. This effect has been already discussed by several previous works and attributed to the fact that for larger heights further out regions of the disc would be illuminated thus contributing more to the observed spectra, which makes it less probable to recover the correct value of the spin \citep[][]{Fabian14, Dovciak2014, Kammoun18}. We note that this effect is seen for both simulations (assuming a radial and constant ionisation profiles). However, when fitted by a model with the constant ionisation profile, the spin parameters are better constrained and consistent with the input values for low heights compared to the former case. This indicates that the ionisation profile of the disc has some non-negligible effect on the spin measurements.

More interestingly, we note that for the two low heights (1.5 and 3~$\rm r_g$) the ability of recovering the correct spin value depends on the ionisation of the disc (see Fig.\,\ref{Par_fig}). The measured spin values are lower for high ionisation states. This can be mainly due to the fact that for high ionisations the reflection features are smoothed out giving a power-law-like spectrum \citep[see Fig. 3 in][for example]{Garcia2013}. Therefore, we suggest that this may be caused by a degeneracy between the reflection and the power-law spectral components which may appear at such high ionisation. As the result, the fitting procedure preferentially underestimates the reflection component. When the over-ionised reflection from the innermost region is not any more modelled as reflection, the fit will then consider that the `observed' reflection is emitted from further out regions of the accretion disc, thus requiring larger values of the ISCO (as compared to the true one). This consequently leads to smaller values of the spin estimates. We note that this applies only for the cases where the spectra were simulated assuming a radially-structured disc ionisation. While the cases assuming a constant ionisation result in accurate spin measurements (see Fig.\,\ref{Par_fig}). This emphasizes the importance of considering radial profiles of ionisation with respect to constant values, if this is the intrinsic profile. Otherwise, the spin measurements can be affected by the above-mentioned spectral model degeneracy.

\section{Discussion}\label{discussion}

In this work, we have studied the effect of a possible radial structure of the disc ionisation on the other parameters of the relativistic reflection models, namely the emissivity index and the spin. The motivation for our work originates from the observationally suggested compactness of the disc-illuminating X-ray source. The innermost regions of the disc are illuminated by much stronger radiation than the outer parts of the disc, which naturally leads to the radial decrease of the disc ionisation for any reasonable density profile of the disc, as illustrated by \cite{2012A&A...545A.106S} (see also Fig.~\ref{ionisation_fig} in the current work). The authors also showed that this radial structure in ionisation may significantly affect the measured radial-emissivity parameter of the reflection model. However, this was shown only for a single and rather extreme example.

Here, we significantly extend the work presented in \cite{2012A&A...545A.106S} by performing a systematic analysis to explore a large range of the parameter space possessing values that allow for a significant detection of the relativistic reflection features with a reasonable exposure time and we study a possible effect of the disc density. 
We investigated different ionisation profiles through assumptions of the disc densities ranging between 10$^{15}$ and 10$^{19}$ cm$^{-3}$, assuming inner ionisation parameters ($\log \xi_{\rm in}$ at the ISCO) in the range 0--4.5. For each ionisation profile of the disc, we performed simulations of the corona being at different heights, ranging between 1.5 and 12~$\rm r_g$ above the black hole, assuming a total observed flux range between $10^{-12}$ and $10^{-10}~ \rm erg~ s^{-1}~ cm^{-2}$, in the 2--10 keV band. First, we simulated spectra assuming a radial profile of the disc ionisation and fitted the spectra with models assuming the constant ionisation. Then, in order to investigate the role of the geometry in giving the steep emissivity profiles, compared to the radial ionisation profiles, we simulated spectra assuming constant disc ionisations and fitted them using models assuming the same criterion. We found that the theoretically expected ionisation gradient in the accretion disc leads to significantly steeper emissivity indices than simply the geometrical effects due to the lamp-post geometry (cf. coloured and grey curves in Fig.~\ref{Par_fig}).

\begin{figure}
\centering
\includegraphics[width=0.9\linewidth]{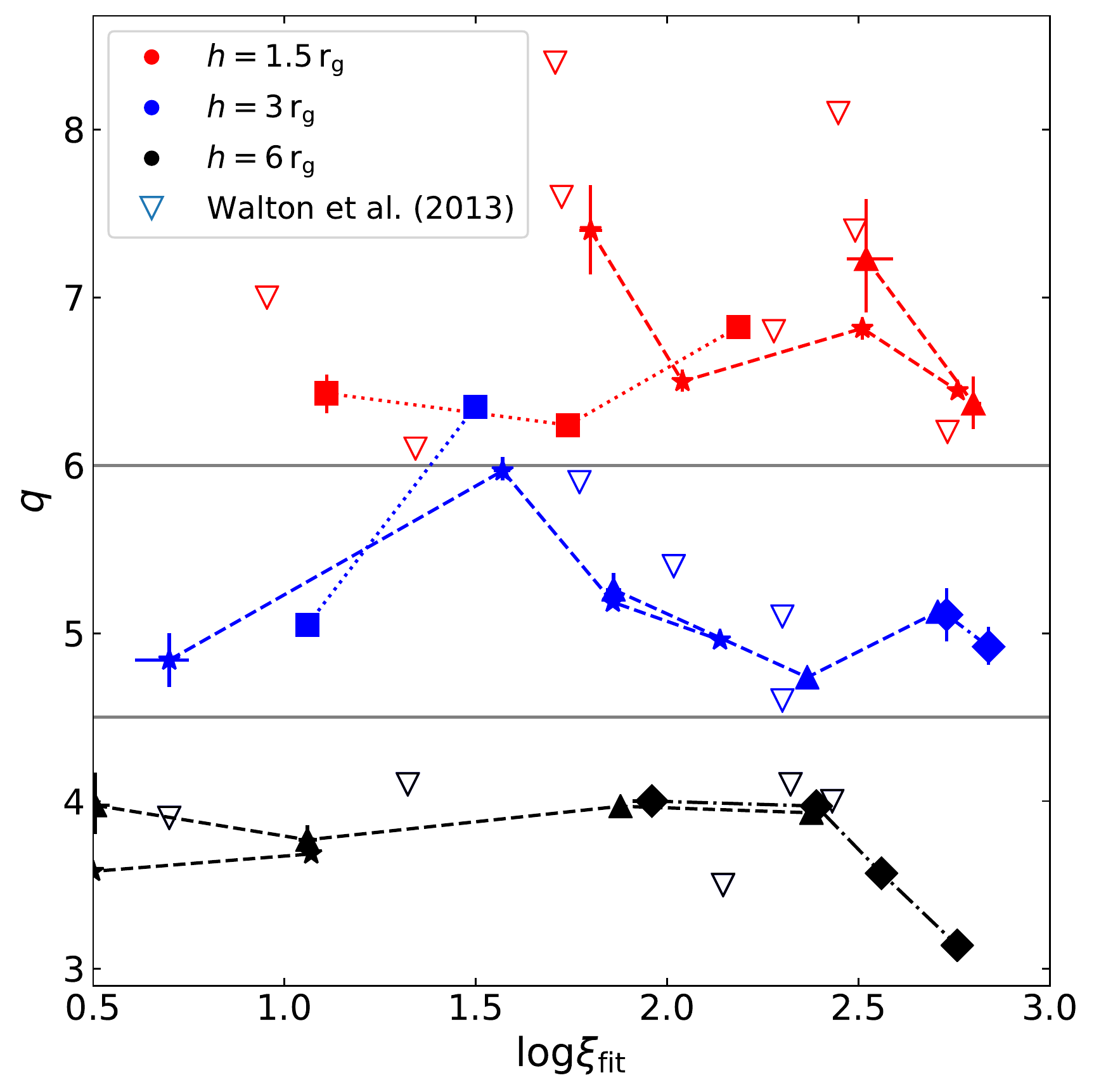}
\caption{Best-fit emissivity indices $q$ as function of the best-fit ionisation parameter $\log \xi_{\rm fit}$, obtained from our simulations for lamp-post heights $h = 1.5,\, 3$ and $6\, \rm r_g$(black, blue and red filled symbols, respectively). The various symbols correspond to different disc densities (using the same code as in Fig. \ref{Par_fig}). The open triangles correspond to the measured values obtained from \citet{Walton2013}, where different colors correspond to different `q-bands' (black: $q \leq 4.5$, blue: $4.5 < q \leq 6 $ and red: $ q>6 $; see text for details).}
\label{fig:walton}
\end{figure}

We stress that we do not present in this work any observational evidence of the detection of an ionisation gradient profile in AGN accretion disc. Instead, our argument is based on simple physical implications of the lamp-post geometry configuration. Fig. \ref{fig:walton} presents a \textit{qualitative} comparison of our results to observational measures. The open symbols in this figure represent the measured emsissivity index versus the ionisation parameter from Table 2 in \citet{Walton2013}. We plot only the cases where  at least a lower limit could be estimated for the emissivity index. We note that, interestingly, none of the observed ionisation parameters in this sample correspond to neither a highly ionised nor completely neutral accretion discs. For the sake of comparison and simplicity, we have divided the observed data points into 3 emissivity-index bands: $q \leq 4.5$ (black open symbols), $4.5 < q \leq 6 $ (blue open symbols) and $ q>6 $ (red open symbols). The filled symbols in Fig. \ref{fig:walton} correspond to our simulations presented in Fig. \ref{Par_fig}. We limit ourselves to the cases where we estimate $\log \xi_{\rm fit}$ to be in the $0.5-3$ range, which broadly corresponds to the observed range for the sample in \citet{Walton2013}. We show the cases which were simulated assuming $h = 1.5$, 3 and 6 $\rm r_g$ (black, blue and red filled symbols), for the various disc densities. This figure shows a fair (qualitative) agreement between the observed data and the results of our simulations which were initially modeled assuming a lamp-post with a self-consistently calculated ionisation gradient of the disc. We note that many uncertainties may affect this comparison mainly whether the reflection strength/fraction would be consistent using the various configurations and models, in addition to the effects of the spin parameter. We note that further work will be needed to assess this issue, ideally requiring the (re-)analysis of the spectra within the context of a gradient ionisation profile, which is beyond the scope of the current work. 

\begin{figure}
\centering
\includegraphics[width=0.9\linewidth]{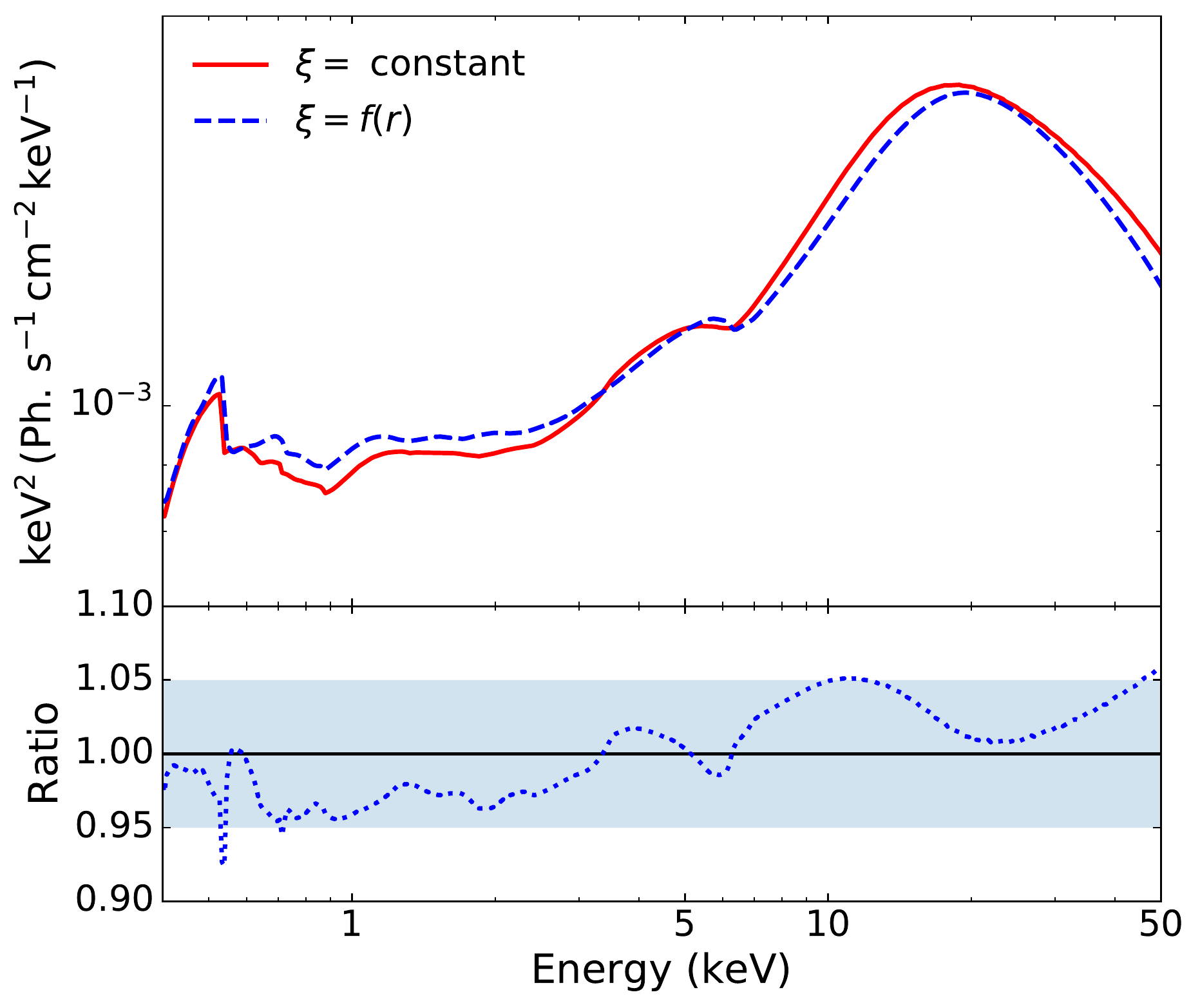}
\caption{Input model used to simulate the data assuming an ionisation gradient in the accretion disc with $h = 3~\rm r_g$, $\log \xi_{\rm in} = 3$ and $n_{\rm H} = 10^{17}~\rm cm^{-3}$ (dashed blue  line) used to simulate the spectrum and the best-fit model obtained by fitting the spectrum in the soft X-rays assuming a constant ionisation (solid red line). We extrapolated both spectra to the hard X-rays in order to check whether the consistency remains at these energies. Bottom panel: The ratio between the two models (bottom panel) is within $\sim 5\%$ for the broad X-ray band.}
\label{fig:hardXray}
\end{figure}

We note that while the consideration of the radial structure in the accretion-disc ionisation is one step towards a more self-consistent description of X-ray reflection spectra, there are several other uncertainties that are mainly caused by our poor knowledge of the exact geometry of the system.
\begin{enumerate}
	
	\item  An extended corona: we assumed here a lamp-post geometry of the corona for simplicity. However, a more realistic configuration is an extended corona \citep{Wilkins2012, Dovciak2016, Tamborra2018, Zhang2018} that would result in a somewhat different illumination of the disc and thus to different ionisation profile. However, models of irradiation by such an extended corona are not yet available for spectral fitting and are dependent on the physical characteristics and structure of such coron\ae.
	
	\item An exact description of the accretion-disc density: we considered here a radial ionisation gradient that is caused by the illumination of the disc with a constant density. Instead, a variation in the density profile of the disc is expected from theoretical models or GRMHD simulations \citep{1973A&A....24..337S, 1973blho.conf..343N, Penna2012}. The exact ionisation profile can be affected by the exact description of the disc density. However, the steep radial decrease of the illumination due to the growing distance from the source can hardly be compensated by a similarly steep increase of the density with radius. Our results are thus qualitatively independent of the accurate density description.

	\item Variability: the primary emission is expected to be variable with time in flux and spectral shape. This would result in a variability in the ionisation profile and in a further complexity in estimating the observed spectrum especially when estimating the response of the accretion disc to any variation in the primary source.  We do not consider these effects in this work.
	
	\item Spin: we have limited our simulations to sources with a spin parameter of 0.94. Lower/higher spin values would result in larger/smaller values of the ISCO which will lead to different ionisation profiles. These configurations are not considered in the current work but should qualitatively provide similar results.
\end{enumerate}
\noindent
It is worth mentioning that our analysis is restricted to the soft X-ray band (below 10~keV). We performed a preliminary test, in order to check whether our results could be affected by including data at higher energies. We consider the case for $h=3~\rm r_g$, $\log \xi_{\rm in} = 3$ and $n_{\rm H} = 10^{17}~\rm cm^{-3}$ in which the spectrum could be fitted by assuming a constant ionisation parameter. In Fig. \ref{fig:hardXray} we show the input spectrum that was used to simulate the data and the best-fit model and we extrapolated both to the hard X-rays. The figure shows that a small difference (less than 5~\%) is expected between the two models even at hard X-rays. We have tested the same for other configurations and they showed similar qualitative and quantitative behaviour. However, a more detailed analysis on how the various parameters might affect the hard X-ray spectra is left for future investigations.

Finally, we stress that our work is not discussing current ways of spin measurements using the constant ionisation in the reflection model. In this work, we show, using high-sensitivity data with future planned instruments, that it is not easy to distinguish between a model with the ionisation gradient and a model with the constant ionisation and the steep radial emissivity for any set of parameters. Most fits with the constant ionisation resulted in C-statistics $C/dof \lesssim 1.5$ (see bottom panel of Fig.~\ref{Par_fig}), thus providing with rather acceptable fits.
However, we consider our work to be useful for understanding how the fitted parameters can be affected by our simplified model assumptions since the disc ionisation gradient is a naturally expected consequence of the lamp-post geometry.
Nevertheless, we plan to extend our work by applying our suggested model to the archival observational data from current instruments such as \textit{XMM-Newton} and compare the results with those obtained here with simulated data for future X-ray mission Athena.

\section{Summary}\label{summary}

X-ray reflection spectra in AGNs and XRBs are thought to originate thanks to the illumination of the accretion disc by a corona of hot electrons in the vicinity of the black hole. For a compact X-ray source located on the black hole rotational axis, this illumination will naturally decrease with radius, implying a radially-structured ionisation of the disc. We showed that such a decrease in ionisation will result in measurements of steep radial emissivities ($q > 5$) assuming a model where the disc ionisation is constant, and fitted as an independent parameter. Steep emissivities are indeed measured in several sources, which may indicate the relevance of the combined effect of the lamp-post geometry and the ionisation gradient. We note that many of these sources show a mild ionisation state, when fitted with a constant-ionisation model, which along with the steep emissivity profiles could indicate the presence of a radial ionisation gradient in the disc. Furthermore, we showed that the ionisation gradient may affect the spin measurements as well. For highly ionised discs, the reflection features are smoothed out, which may result in inaccurate spin measurements if one assumes a constant ionisation (this may underestimate the real contribution by reflection to the total spectrum). 
We also confirmed previous results stating that spin measurements would not be accurate if the corona is located at large heights. Finally, we showed that the level of the disc density does not play any important role in determining the emissivity profile.	

\section*{Acknowledgements}
We would like to thank the referee, Dr. Javier Garc\'{i}a, for useful comments and suggestions which helped improving the clarity of our manuscript.  This work has been started during a visit, under the ERASMUS+ Programme - Student Mobility for Traineeship-Project KTEU-ET (Key to Europe Erasmus Traineeship), of ESK to the Astronomical Insitute  of the Czech Academy of Sciences, whose hospitality he gratefully acknowledges. JS and MD acknowledge financial support from the Grant Agency of the Czech Republic within the project No. 17-02430S and MEYS project LTAUSA17095.

\bibliographystyle{mnras}
\bibliography{references} 

\begin{thebibliography}{}
\makeatletter
\relax
\def\mn@urlcharsother{\let\do\@makeother \do\$\do\&\do\#\do\^\do\_\do\%\do\~}
\def\mn@doi{\begingroup\mn@urlcharsother \@ifnextchar [ {\mn@doi@}
  {\mn@doi@[]}}
\def\mn@doi@[#1]#2{\def\@tempa{#1}\ifx\@tempa\@empty \href
  {http://dx.doi.org/#2} {doi:#2}\else \href {http://dx.doi.org/#2} {#1}\fi
  \endgroup}
\def\mn@eprint#1#2{\mn@eprint@#1:#2::\@nil}
\def\mn@eprint@arXiv#1{\href {http://arxiv.org/abs/#1} {{\tt arXiv:#1}}}
\def\mn@eprint@dblp#1{\href {http://dblp.uni-trier.de/rec/bibtex/#1.xml}
  {dblp:#1}}
\def\mn@eprint@#1:#2:#3:#4\@nil{\def\@tempa {#1}\def\@tempb {#2}\def\@tempc
  {#3}\ifx \@tempc \@empty \let \@tempc \@tempb \let \@tempb \@tempa \fi \ifx
  \@tempb \@empty \def\@tempb {arXiv}\fi \@ifundefined
  {mn@eprint@\@tempb}{\@tempb:\@tempc}{\expandafter \expandafter \csname
  mn@eprint@\@tempb\endcsname \expandafter{\@tempc}}}

\bibitem[\protect\citeauthoryear{Arnaud}{Arnaud}{1996}]{Arnaud1996}
Arnaud K.~A.,  1996, in Astronomical Data Analysis Software and Systems V.
  p.~17, \url {http://adsabs.harvard.edu/abs/1996ASPC..101...17A}

\bibitem[\protect\citeauthoryear{Ballantyne, Ross  \& Fabian}{Ballantyne
  et~al.}{2001}]{Ballantyne2001}
Ballantyne D.,  Ross R.,   Fabian A.~C.,  2001, \mn@doi [\mnras]
  {10.1046/j.1365-8711.2001.04432.x}, 327, 10

\bibitem[\protect\citeauthoryear{Beckwith \& Done}{Beckwith \&
  Done}{2004}]{2004MNRAS.352..353B}
Beckwith K.,  Done C.,  2004, \mn@doi [\mnras]
  {10.1111/j.1365-2966.2004.07955.x}, 352, 353

\bibitem[\protect\citeauthoryear{Brandt \& Alexander}{Brandt \&
  Alexander}{2015}]{Brandt2015}
Brandt W.~N.,  Alexander D.~M.,  2015, \mn@doi [\aapr]
  {10.1007/s00159-014-0081-z}, 23, 93

\bibitem[\protect\citeauthoryear{Brenneman \& Reynolds}{Brenneman \&
  Reynolds}{2006}]{2006ApJ...652.1028B}
Brenneman L.,  Reynolds C.,  2006, \mn@doi [\apj] {10.1086/508146}, 652, 1028

\bibitem[\protect\citeauthoryear{{Cash}}{{Cash}}{1976}]{Cash76}
{Cash} W.,  1976, \aap, \href
  {http://adsabs.harvard.edu/abs/1976A%26A....52..307C} {52, 307}

\bibitem[\protect\citeauthoryear{{Chartas}, {Kochanek}, {Dai}, {Poindexter}  \&
  {Garmire}}{{Chartas} et~al.}{2009}]{Chartas09}
{Chartas} G.,  {Kochanek} C.~S.,  {Dai} X.,  {Poindexter} S.,   {Garmire} G.,
  2009, \mn@doi [\apj] {10.1088/0004-637X/693/1/174}, \href
  {http://adsabs.harvard.edu/abs/2009ApJ...693..174C} {693, 174}

\bibitem[\protect\citeauthoryear{Dauser, Wilms, Reynolds  \& Brenneman}{Dauser
  et~al.}{2010}]{Dauser2010}
Dauser T.,  Wilms J.,  Reynolds C.~S.,   Brenneman L.~W.,  2010, \mn@doi
  [\mnras] {10.1111/j.1365-2966.2010.17393.x}, 409, 1534

\bibitem[\protect\citeauthoryear{Dauser et~al.,}{Dauser
  et~al.}{2012}]{2012MNRAS.422.1914D}
Dauser T.,  et~al., 2012, \mn@doi [\mnras] {10.1111/j.1365-2966.2011.20356.x},
  422, 1914

\bibitem[\protect\citeauthoryear{Dauser, Garc{\'{i}}a, Wilms, Bock, Brenneman,
  Falanga, Fukumura  \& Reynolds}{Dauser et~al.}{2013}]{Dauser2013}
Dauser T.,  Garc{\'{i}}a J.,  Wilms J.,  Bock M.,  Brenneman L.~W.,  Falanga
  M.,  Fukumura K.,   Reynolds C.,  2013, \mn@doi [\mnras]
  {10.1093/mnras/sts710}, 430, 1694

\bibitem[\protect\citeauthoryear{Dauser, Garc{\'{i}}a, Parker, Fabian  \&
  Wilms}{Dauser et~al.}{2014}]{Dauser2014}
Dauser T.,  Garc{\'{i}}a J.,  Parker M.~L.,  Fabian A.~C.,   Wilms J.,  2014,
  \mn@doi [\mnras] {10.1093/mnrasl/slu125}, 444, L100

\bibitem[\protect\citeauthoryear{Dov{\v{c}}iak \& Done}{Dov{\v{c}}iak \&
  Done}{2016}]{Dovciak2016}
Dov{\v{c}}iak M.,  Done C.,  2016, \mn@doi [Astronomische Nachrichten]
  {10.1002/asna.201612327}, 337, 441

\bibitem[\protect\citeauthoryear{Dov{\v{c}}iak, Karas, Martocchia, Matt  \&
  Yaqoob}{Dov{\v{c}}iak et~al.}{2004a}]{Dovciak2004}
Dov{\v{c}}iak M.,  Karas V.,  Martocchia A.,  Matt G.,   Yaqoob T.,  2004a, in
  Processes in the vicinity of black holes and neutron stars.  (\mn@eprint
  {arXiv} {0407330}), \url {http://arxiv.org/abs/astro-ph/0407330}

\bibitem[\protect\citeauthoryear{Dov{\v{c}}iak, Karas  \& Yaqoob}{Dov{\v{c}}iak
  et~al.}{2004b}]{ky}
Dov{\v{c}}iak M.,  Karas V.,   Yaqoob T.,  2004b, \mn@doi [\apjs]
  {10.1086/421115}, 153, 205

\bibitem[\protect\citeauthoryear{Dov{\v{c}}iak, Muleri, Goosmann, Karas  \&
  Matt}{Dov{\v{c}}iak et~al.}{2011}]{2011ApJ...731...75D}
Dov{\v{c}}iak M.,  Muleri F.,  Goosmann R.,  Karas V.,   Matt G.,  2011,
  \mn@doi [\apj] {10.1088/0004-637X/731/1/75}, 731, 75

\bibitem[\protect\citeauthoryear{Dov{\v{c}}iak, Svoboda, Goosmann, Karas, Matt
  \& Sochora}{Dov{\v{c}}iak et~al.}{2014}]{Dovciak2014}
Dov{\v{c}}iak M.,  Svoboda J.,  Goosmann R.~W.,  Karas V.,  Matt G.,   Sochora
  V.,  2014, eprint arXiv:1412.8627

\bibitem[\protect\citeauthoryear{{Fabian} et~al.,}{{Fabian}
  et~al.}{2012a}]{Fabian20121H}
{Fabian} A.~C.,  et~al., 2012a, \mn@doi [\mnras]
  {10.1111/j.1365-2966.2011.19676.x}, \href
  {http://adsabs.harvard.edu/abs/2012MNRAS.419..116F} {419, 116}

\bibitem[\protect\citeauthoryear{Fabian et~al.,}{Fabian
  et~al.}{2012b}]{2012MNRAS.424..217F}
Fabian A.~C.,  et~al., 2012b, \mn@doi [\mnras]
  {10.1111/j.1365-2966.2012.21185.x}, 424, 217

\bibitem[\protect\citeauthoryear{{Fabian}, {Parker}, {Wilkins}, {Miller},
  {Kara}, {Reynolds}  \& {Dauser}}{{Fabian} et~al.}{2014}]{Fabian14}
{Fabian} A.~C.,  {Parker} M.~L.,  {Wilkins} D.~R.,  {Miller} J.~M.,  {Kara} E.,
   {Reynolds} C.~S.,   {Dauser} T.,  2014, \mn@doi [\mnras]
  {10.1093/mnras/stu045}, \href
  {http://adsabs.harvard.edu/abs/2014MNRAS.439.2307F} {439, 2307}

\bibitem[\protect\citeauthoryear{Garc{\'{i}}a, Dauser, Reynolds, Kallman,
  McClintock, Wilms  \& Eikmann}{Garc{\'{i}}a et~al.}{2013}]{Garcia2013}
Garc{\'{i}}a J.,  Dauser T.,  Reynolds C.,  Kallman T.~R.,  McClintock J.~E.,
  Wilms J.,   Eikmann W.,  2013, \mn@doi [\apj] {10.1088/0004-637X/768/2/146},
  768, 146

\bibitem[\protect\citeauthoryear{{Garc{\'{\i}}a}, {Fabian}, {Kallman},
  {Dauser}, {Parker}, {McClintock}, {Steiner}  \& {Wilms}}{{Garc{\'{\i}}a}
  et~al.}{2016}]{Garcia16}
{Garc{\'{\i}}a} J.~A.,  {Fabian} A.~C.,  {Kallman} T.~R.,  {Dauser} T.,
  {Parker} M.~L.,  {McClintock} J.~E.,  {Steiner} J.~F.,   {Wilms} J.,  2016,
  \mn@doi [\mnras] {10.1093/mnras/stw1696}, \href
  {http://adsabs.harvard.edu/abs/2016MNRAS.462..751G} {462, 751}

\bibitem[\protect\citeauthoryear{{George}, {Fabian}, {Nandra}, {Pounds}  \&
  {Stewart}}{{George} et~al.}{1989}]{George89}
{George} I.~M.,  {Fabian} A.~C.,  {Nandra} K.,  {Pounds} K.~A.,   {Stewart}
  G.~C.,  1989, in {Hunt} J.,  {Battrick} B.,  eds,  ESA Special Publication
  Vol. 296, Two Topics in X-Ray Astronomy, Volume 1: X Ray Binaries. Volume 2:
  AGN and the X Ray Background.

\bibitem[\protect\citeauthoryear{{Gonzalez}, {Wilkins}  \& {Gallo}}{{Gonzalez}
  et~al.}{2017}]{Gonzalez17}
{Gonzalez} A.~G.,  {Wilkins} D.~R.,   {Gallo} L.~C.,  2017, \mn@doi [\mnras]
  {10.1093/mnras/stx2080}, \href
  {http://adsabs.harvard.edu/abs/2017MNRAS.472.1932G} {472, 1932}

\bibitem[\protect\citeauthoryear{{Jiang} et~al.,}{{Jiang}
  et~al.}{2018}]{Jiang18}
{Jiang} J.,  et~al., 2018, \mn@doi [\mnras] {10.1093/mnras/sty836}, \href
  {http://adsabs.harvard.edu/abs/2018MNRAS.477.3711J} {477, 3711}

\bibitem[\protect\citeauthoryear{{Kammoun}, {Nardini}  \& {Risaliti}}{{Kammoun}
  et~al.}{2018}]{Kammoun18}
{Kammoun} E.~S.,  {Nardini} E.,   {Risaliti} G.,  2018, \mn@doi [\aap]
  {10.1051/0004-6361/201732377}, \href
  {http://adsabs.harvard.edu/abs/2018A%26A...614A..44K} {614, A44}

\bibitem[\protect\citeauthoryear{{Kosec}, {Buisson}, {Parker}, {Pinto},
  {Fabian}  \& {Walton}}{{Kosec} et~al.}{2018}]{Kosec18}
{Kosec} P.,  {Buisson} D.~J.~K.,  {Parker} M.~L.,  {Pinto} C.,  {Fabian} A.~C.,
    {Walton} D.~J.,  2018, \mn@doi [\mnras] {10.1093/mnras/sty2342}, \href
  {http://adsabs.harvard.edu/abs/2018MNRAS.481..947K} {481, 947}

\bibitem[\protect\citeauthoryear{Marinucci et~al.,}{Marinucci
  et~al.}{2014a}]{Marinucci2014}
Marinucci A.,  et~al., 2014a, \mn@doi [\mnras] {10.1093/mnras/stu404}, 440,
  2347

\bibitem[\protect\citeauthoryear{Marinucci et~al.,}{Marinucci
  et~al.}{2014b}]{Marinucci2014a}
Marinucci A.,  et~al., 2014b, \mn@doi [\apj] {10.1088/0004-637X/787/1/83}, 787,
  83

\bibitem[\protect\citeauthoryear{{Martocchia} \& {Matt}}{{Martocchia} \&
  {Matt}}{1996}]{Martocchia96}
{Martocchia} A.,  {Matt} G.,  1996, \mn@doi [\mnras] {10.1093/mnras/282.4.L53},
  \href {http://adsabs.harvard.edu/abs/1996MNRAS.282L..53M} {282, L53}

\bibitem[\protect\citeauthoryear{{Martocchia}, {Karas}  \& {Matt}}{{Martocchia}
  et~al.}{2000}]{Martocchia00}
{Martocchia} A.,  {Karas} V.,   {Matt} G.,  2000, \mn@doi [\mnras]
  {10.1046/j.1365-8711.2000.03205.x}, \href
  {http://adsabs.harvard.edu/abs/2000MNRAS.312..817M} {312, 817}

\bibitem[\protect\citeauthoryear{{Martocchia}, {Matt}  \& {Karas}}{{Martocchia}
  et~al.}{2002}]{Martocchia02}
{Martocchia} A.,  {Matt} G.,   {Karas} V.,  2002, \mn@doi [\aap]
  {10.1051/0004-6361:20020089}, \href
  {http://adsabs.harvard.edu/abs/2002A%26A...383L..23M} {383, L23}

\bibitem[\protect\citeauthoryear{Matt, Perola  \& Piro}{Matt
  et~al.}{1991}]{Matt1991}
Matt G.,  Perola G.~C.,   Piro L.,  1991, Astronomy and Astrophysics, 247, 25

\bibitem[\protect\citeauthoryear{Merloni}{Merloni}{2016}]{Merloni2016}
Merloni A.,  2016, \mn@doi [Lecture Notes in Physics]
  {10.1007/978-3-319-19416-5_4}, 905, 101

\bibitem[\protect\citeauthoryear{Miller}{Miller}{2007}]{2007ARA&A..45..441M}
Miller J.,  2007, \mn@doi [\araa] {10.1146/annurev.astro.45.051806.110555}, 45,
  441

\bibitem[\protect\citeauthoryear{{Miller} et~al.,}{{Miller}
  et~al.}{2013}]{Miller2013}
{Miller} J.~M.,  et~al., 2013, \mn@doi [\apjl] {10.1088/2041-8205/775/2/L45},
  \href {http://adsabs.harvard.edu/abs/2013ApJ...775L..45M} {775, L45}

\bibitem[\protect\citeauthoryear{{Miller} et~al.,}{{Miller}
  et~al.}{2018}]{2018ApJ...860L..28M}
{Miller} J.~M.,  et~al., 2018, \mn@doi [\apjl] {10.3847/2041-8213/aacc61},
  \href {http://adsabs.harvard.edu/abs/2018ApJ...860L..28M} {860, L28}

\bibitem[\protect\citeauthoryear{Miniutti \& Fabian}{Miniutti \&
  Fabian}{2004}]{Miniutti2004}
Miniutti G.,  Fabian A.~C.,  2004, \mn@doi [\mnras]
  {10.1111/j.1365-2966.2004.07611.x}, 349, 1435

\bibitem[\protect\citeauthoryear{Miniutti, Fabian  \& Miller}{Miniutti
  et~al.}{2004}]{2004MNRAS.351..466M}
Miniutti G.,  Fabian A.~C.,   Miller J.,  2004, \mn@doi [\mnras]
  {10.1111/j.1365-2966.2004.07794.x}, 351, 466

\bibitem[\protect\citeauthoryear{{Mosquera}, {Kochanek}, {Chen}, {Dai},
  {Blackburne}  \& {Chartas}}{{Mosquera} et~al.}{2013}]{Mosquera13}
{Mosquera} A.~M.,  {Kochanek} C.~S.,  {Chen} B.,  {Dai} X.,  {Blackburne}
  J.~A.,   {Chartas} G.,  2013, \mn@doi [\apj] {10.1088/0004-637X/769/1/53},
  \href {http://adsabs.harvard.edu/abs/2013ApJ...769...53M} {769, 53}

\bibitem[\protect\citeauthoryear{Nandra et~al.,}{Nandra
  et~al.}{2013}]{Nandra2013}
Nandra K.,  et~al., 2013, arXiv:1306.2307

\bibitem[\protect\citeauthoryear{Nayakshin, Kazanas  \& Kallman}{Nayakshin
  et~al.}{2000}]{Nayakshin2000}
Nayakshin S.,  Kazanas D.,   Kallman T.,  2000, \mn@doi [\apj]
  {10.1086/309054}, 537, 833

\bibitem[\protect\citeauthoryear{Nied{\'{z}}wiecki \&
  Miyakawa}{Nied{\'{z}}wiecki \& Miyakawa}{2010}]{Niedzwiecki2010}
Nied{\'{z}}wiecki A.,  Miyakawa T.,  2010, \mn@doi [\aap]
  {10.1051/0004-6361/200911919}, 509, A22

\bibitem[\protect\citeauthoryear{Nied{\'{z}}wiecki \&
  {\.{Z}}ycki}{Nied{\'{z}}wiecki \& {\.{Z}}ycki}{2008}]{Niedzwiecki2008}
Nied{\'{z}}wiecki A.,  {\.{Z}}ycki P.~T.,  2008, \mn@doi [\mnras]
  {10.1111/j.1365-2966.2008.12735.x}, 386, 759

\bibitem[\protect\citeauthoryear{Nied{\'{z}}wiecki, Zdziarski  \&
  Szanecki}{Nied{\'{z}}wiecki et~al.}{2016}]{Nied2016}
Nied{\'{z}}wiecki A.,  Zdziarski A.~A.,   Szanecki M.,  2016, \mn@doi [\apj]
  {10.3847/2041-8205/821/1/L1}, 821, L1

\bibitem[\protect\citeauthoryear{Novikov \& Thorne}{Novikov \&
  Thorne}{1973}]{1973blho.conf..343N}
Novikov I.,  Thorne K.,  1973, in Black Holes (Les Astres Occlus). pp 343--450

\bibitem[\protect\citeauthoryear{Parker et~al.,}{Parker
  et~al.}{2014}]{Parker2014}
Parker M.~L.,  et~al., 2014, \mn@doi [\mnras] {10.1093/mnras/stu1246}, 443,
  1723

\bibitem[\protect\citeauthoryear{{Penna}, {S{\c a}owski}  \&
  {McKinney}}{{Penna} et~al.}{2012}]{Penna2012}
{Penna} R.~F.,  {S{\c a}owski} A.,   {McKinney} J.~C.,  2012, \mn@doi [\mnras]
  {10.1111/j.1365-2966.2011.20084.x}, \href
  {http://adsabs.harvard.edu/abs/2012MNRAS.420..684P} {420, 684}

\bibitem[\protect\citeauthoryear{Ponti et~al.,}{Ponti
  et~al.}{2010}]{2010MNRAS.406.2591P}
Ponti G.,  et~al., 2010, \mn@doi [\mnras] {10.1111/j.1365-2966.2010.16852.x},
  406, 2591

\bibitem[\protect\citeauthoryear{{Ravera} et~al.,}{{Ravera}
  et~al.}{2014}]{Ravera14}
{Ravera} L.,  et~al., 2014, in Space Telescopes and Instrumentation 2014:
  Ultraviolet to Gamma Ray. p. 91442L, \mn@doi{10.1117/12.2055884}

\bibitem[\protect\citeauthoryear{{Reis} \& {Miller}}{{Reis} \&
  {Miller}}{2013}]{Reis13}
{Reis} R.~C.,  {Miller} J.~M.,  2013, \mn@doi [\apjl]
  {10.1088/2041-8205/769/1/L7}, \href
  {http://adsabs.harvard.edu/abs/2013ApJ...769L...7R} {769, L7}

\bibitem[\protect\citeauthoryear{Reynolds}{Reynolds}{2013}]{Reynolds2013}
Reynolds C.,  2013, Space Science Reviews, 183, 277

\bibitem[\protect\citeauthoryear{Reynolds \& Begelman}{Reynolds \&
  Begelman}{1997}]{1997ApJ...488..109R}
Reynolds C.,  Begelman M.,  1997, \mn@doi [\apj] {10.1086/304703}, 488, 109

\bibitem[\protect\citeauthoryear{Shakura \& Sunyaev}{Shakura \&
  Sunyaev}{1973}]{1973A&A....24..337S}
Shakura N.,  Sunyaev R.,  1973, \aap, 24, 337

\bibitem[\protect\citeauthoryear{Svoboda, Dov{\v{c}}iak, Goosmann  \&
  Karas}{Svoboda et~al.}{2009}]{Svoboda2009}
Svoboda J.,  Dov{\v{c}}iak M.,  Goosmann R.~W.,   Karas V.,  2009, \mn@doi
  [\aap] {10.1051/0004-6361/200911941}, 507, 18

\bibitem[\protect\citeauthoryear{Svoboda, Dov{\v{c}}iak, Goosmann, Jethwa,
  Karas, Miniutti  \& Guainazzi}{Svoboda et~al.}{2012}]{2012A&A...545A.106S}
Svoboda J.,  Dov{\v{c}}iak M.,  Goosmann R.,  Jethwa P.,  Karas V.,  Miniutti
  G.,   Guainazzi M.,  2012, \mn@doi [\aap] {10.1051/0004-6361/201219701}, 545,
  A106

\bibitem[\protect\citeauthoryear{Svoboda, Beuchert, Guainazzi, Longinotti,
  Piconcelli  \& Wilms}{Svoboda et~al.}{2015}]{Svoboda2015}
Svoboda J.,  Beuchert T.,  Guainazzi M.,  Longinotti A.~L.,  Piconcelli E.,
  Wilms J.,  2015, \mn@doi [\aap] {10.1051/0004-6361/201425453}, 578, A96

\bibitem[\protect\citeauthoryear{{Tamborra}, {Matt}, {Bianchi}  \& {Dov{\v
  c}iak}}{{Tamborra} et~al.}{2018}]{Tamborra2018}
{Tamborra} F.,  {Matt} G.,  {Bianchi} S.,   {Dov{\v c}iak} M.,  2018, \mn@doi
  [\aap] {10.1051/0004-6361/201732023}, \href
  {http://adsabs.harvard.edu/abs/2018A%26A...619A.105T} {619, A105}

\bibitem[\protect\citeauthoryear{Tanaka et~al.,}{Tanaka
  et~al.}{1995}]{1995Natur.375..659T}
Tanaka Y.,  et~al., 1995, \mn@doi [\nat] {10.1038/375659a0}, 375, 659

\bibitem[\protect\citeauthoryear{{Tomsick} et~al.,}{{Tomsick}
  et~al.}{2014}]{Tomsick2014}
{Tomsick} J.~A.,  et~al., 2014, \mn@doi [\apj] {10.1088/0004-637X/780/1/78},
  \href {http://adsabs.harvard.edu/abs/2014ApJ...780...78T} {780, 78}

\bibitem[\protect\citeauthoryear{Walton, Nardini, Fabian, Gallo  \&
  Reis}{Walton et~al.}{2012}]{Walton2012}
Walton D.~J.,  Nardini E.,  Fabian A.~C.,  Gallo L.~C.,   Reis R.~C.,  2012,
  \mn@doi [\mnras] {10.1093/mnras/sts227}, 428, 2901

\bibitem[\protect\citeauthoryear{{Walton}, {Nardini}, {Fabian}, {Gallo}  \&
  {Reis}}{{Walton} et~al.}{2013}]{Walton2013}
{Walton} D.~J.,  {Nardini} E.,  {Fabian} A.~C.,  {Gallo} L.~C.,   {Reis} R.~C.,
   2013, \mn@doi [\mnras] {10.1093/mnras/sts227}, \href
  {http://adsabs.harvard.edu/abs/2013MNRAS.428.2901W} {428, 2901}

\bibitem[\protect\citeauthoryear{Wilkins \& Fabian}{Wilkins \&
  Fabian}{2011}]{Wilkins2011}
Wilkins D.~R.,  Fabian A.~C.,  2011, \mn@doi [\mnras]
  {10.1111/j.1365-2966.2011.18458.x}, 414, 1269

\bibitem[\protect\citeauthoryear{Wilkins \& Fabian}{Wilkins \&
  Fabian}{2012}]{Wilkins2012}
Wilkins D.~R.,  Fabian A.~C.,  2012, \mn@doi [Monthly Notices of the Royal
  Astronomical Society] {10.1111/j.1365-2966.2012.21308.x}, 424, 1284

\bibitem[\protect\citeauthoryear{Wilkins \& Gallo}{Wilkins \&
  Gallo}{2015}]{Wilkins2015}
Wilkins D.~R.,  Gallo L.~C.,  2015, \mn@doi [\mnras] {10.1093/mnras/stv162},
  449, 129

\bibitem[\protect\citeauthoryear{{Wilkins}, {Kara}, {Fabian}  \&
  {Gallo}}{{Wilkins} et~al.}{2014}]{Wilkins14}
{Wilkins} D.~R.,  {Kara} E.,  {Fabian} A.~C.,   {Gallo} L.~C.,  2014, \mn@doi
  [\mnras] {10.1093/mnras/stu1273}, \href
  {http://adsabs.harvard.edu/abs/2014MNRAS.443.2746W} {443, 2746}

\bibitem[\protect\citeauthoryear{{Zhang}}{{Zhang}}{2018}]{Zhang2018}
{Zhang} W.,  2018, in 42nd COSPAR Scientific Assembly. pp E1.4--52--18

\bibitem[\protect\citeauthoryear{Zycki, Krolik, Zdziarski  \& Kallman}{Zycki
  et~al.}{1994}]{Zycki1994}
Zycki P.,  Krolik J.,  Zdziarski A.,   Kallman T.,  1994, \mn@doi [\apj]
  {10.1086/175024}, 437, 597

\makeatother
\end{thebibliography}


\bsp	
\label{lastpage}
\end{document}